\documentclass[12pt]{article}

\title{Monetary Measures of Risk\footnote{This paper was written in 2015 as a contribution to Wiley Encyclopedia of Operations Research and Management Science. It is accepted and will appear in the second edition of EORMS.}}

\author{Andreas H. Hamel\footnote{Free University of Bozen-Bolzano, Faculty of Economics and Management, \href{mailto:andreas.hamel@unibz.it}{andreas.hamel@unibz.it}}}
\date{}

\usepackage{array} % for better arrays (eg matrices) in maths
\usepackage{verbatim} % adds environment for commenting out blocks of text & for better verbatim
\usepackage{amsmath, amssymb, enumerate, color}
\usepackage{stmaryrd}

\newtheorem{theorem}{Theorem}

\newtheorem{definition}[theorem]{Definition}
\newtheorem{proposition}[theorem]{Proposition}

\numberwithin{figure}{section}    % Abbildungen mit f¸hrender Kapitelnummer
\numberwithin{table}{section}     % Tabellen mit f¸hrender Kapitelnummer
\newcommand{\of}[1]{\ensuremath{\left( #1 \right)}}
\newcommand{\norm}[1]{\ensuremath{\left\| #1 \right\|}}
\newcommand{\abs}[1]{\ensuremath{\left| #1 \right|}}
\newcommand{\cb}[1]{\ensuremath{ \left\{ #1 \right\} }}
\newcommand{\sqb}[1]{\ensuremath{ \left[ #1 \right] }}

\newcommand{\eps}{\ensuremath{\varepsilon}}

\newcommand{\F}{\ensuremath{\mathcal{F}}}

\DeclareMathOperator*{\essinf}{essinf}
\DeclareMathOperator*{\esssup}{esssup}

\newcommand{\R}{\mathrm{I\negthinspace R}}

\newcommand{\Int}{{\rm int\,}}

\newcommand{\One}{\mathrm{1\negthickspace I}}

\usepackage[%backref=page,
colorlinks=true, linkcolor=blue, citecolor=blue, urlcolor=blue, filecolor=blue
auskommentieren und unten pdfborder aktivieren]{hyperref}
\definecolor{color0}{gray}{.50}
\definecolor{color1}{rgb}{0,.2,.8}
\definecolor{color2}{rgb}{1,.2,0}
\definecolor{color3}{rgb}{.8,.5,1}
\newcommand{\f}{\color{color1}}

\begin{document}

\maketitle

A {\em monetary risk measure} is a mathematical tool for quantifying the risk of a random future gain (or loss) which is denoted in (discounted) units of a reference instrument (a currency, for example). As such, it is a real-valued function, and it is convenient to allow for the value $+\infty$. The greater the value of the risk measure, the higher the risk.

Two elementary mathematical properties turned out to be crucial. Both have a straightforward and convincing economic interpretation.

The first one is a monotonicity property: if gain $X$ is not less than gain $Y$ no matter what happens in the world, then the risk of $X$ should not be greater than the risk of $Y$.

The second one is additivity with respect to a riskless reference instrument: if one adds $s$ units of the reference instrument (e.g., cash) to the (discounted) random gain (e.g., as a deposit), then the risk (i.e., the value of the risk measure) decreases by $s$. Because of the immediate interpretation of the value of such risk measures as capital requirements, they are also called {\em monetary measures of risk} \cite{FoellmerSchied11}.

This second property, called {\em cash-additivity}, has remarkable mathematical consequences. Its economic interpretation, `linearity in payments' has been pointed out already very clearly in \cite[p. 101]{Yaari87E} by Yaari, and it became popular through the work \cite{ArtznerEtAl99MF} by Artzner et al. Cash-additivity goes by many names, for example ``translation invariance," (already in \cite{WangYoungPanjer97IME}, also \cite{ArtznerEtAl99MF}), ``translation equivariance" (\cite{RuszczynskiShapiro06MOR}), or just ``additivity" (\cite[p. 1455]{BenTalTeboulle86MS}).

Following the famous \cite{Markowitz52JF}, the variance of a random variable was used as a risk measure in portfolio selection problems (see also {\f 3.2.2.2}\footnote{Such labels refer to other articles in EORMS.} and {\f 3.2.2.4}). However, it is neither monotone, nor cash-additive. Moreover, it weighs (random) gains and losses in a symmetrical way which is not a desirable property of a (financial) risk measure and, on an even deeper level, the variance is not consistent with important stochastic dominance orders as already discussed, e.g., in \cite{OgryczakRuszczinski99EJOR}. On the other hand, in the financial practice the (cash-additive)  so-called Value-at-Risk was (and still is) widely used as a risk measure. Its drawback turned out to be the missing convexity: diversification is not generally supported by the Value-at-Risk. Therefore, a new class of (monotone and cash-additive) risk measures, called {\em coherent}, was introduced in \cite{ArtznerEtAl99MF} (see also {\f 4.6.3.3}).

%%%New section
\section{Risk measures and acceptance sets} 

Let $\of{\Omega, \F, P}$ be a probability space, $p \in [0, \infty]$ and $L^p := L^p(\Omega, \F, P)$ be the linear space of all (equivalence classes of) univariate, to the $p$th power (absolutely) integrable random variables where $X_1 \colon \Omega \to \R$ and $X_2 \colon \Omega \to \R$ generate the same element of $L^p$ whenever $P\of{\cb{\omega \in \Omega \mid X_1(\omega) = X_2(\omega)}} = 1$. If $p=0$, $L^0$ is the space of all random variables. If $p = \infty$, $L^\infty := L^\infty(\Omega, \F, P)$ is the space of essentially bounded random variables. An inequality like  $X_1 \leq X_2$ for two elements of $X_1, X_2 \in L^p$ is understood $P$-almost surely, i.e. $P\of{\cb{\omega \in \Omega \mid X_1(\omega) \leq X_2(\omega)}} = 1$. The element $\One  \in L^p$ denotes the function whose value is 1 $P$-almost surely, and $L^p_+ = \cb{X \in L^p \mid 0 \leq X}$.

A function $\varrho \colon L^p \to \R\cup\cb{+\infty}$ is called {\em monotone} if $X, Y \in L^p$, $X \leq Y$ imply $\varrho(Y) \leq \varrho(X)$, and it is called {\em cash-additive} if
\begin{equation}
\label{EqCashAdd}
\forall X \in  L^p, \; \forall r \in \R \colon \varrho(X + r \One) = \varrho(X) - r.
\end{equation}

\begin{definition}
\label{DefRiskMeasure} A {\em risk measure} is a function $\varrho \colon L^p \to \R\cup\cb{+\infty}$ which is monotone, cash-additive and satisfies $\varrho(0) \in \R$. 
\end{definition}

Risk measures can also be defined on other linear spaces of random variables, see, for example, \cite{CheriditoLi08MFE}. It is remarkable, though mathematically not difficult, that risk measures are basically in one-to-one correspondence with their acceptance sets (see also {\f 3.2.4.1}). This fact depends almost entirely on property \eqref{EqCashAdd}. Here are the necessary concepts. A set $A \subseteq L^p$ is called {\em monotone} if $A + L^p_+ \subseteq A$, and it is called {\em directionally closed} if $X \in A$, $\cb{r_n}_{n=0,1, \ldots} \subseteq \R_+$, $\lim_{n \to \infty} r_n = 0$ and $X + r_n\One \in A$ for all $n = 0,1, \ldots$ imply $X \in A$.

\begin{definition}
\label{DefSccetanceSet} An {\em acceptance set} is a set $A \subseteq L^p$ which is monotone, directionally closed and satisfies $A \cap \R\One  \neq \emptyset$ as well as $(L^p\backslash A) - \R\One = L^p$.
\end{definition}

The correspondence between acceptance sets and risk measures is established in the following result.

\begin{proposition}
\label{PropPrimalRep} If $A \subseteq L^p$ is an acceptance set, then the function $\varrho_A$ on $L^p$ defined by
\begin{equation}
\label{EqRhoA}
\varrho_A(X) = \inf\cb{s \in \R \mid X + s\One \in A}
\end{equation}
is a risk measure. If $\varrho \colon L^p \to \R\cup\cb{+\infty}$ is a risk measure, then the set
\begin{equation}
\label{EqARho}
A_\varrho = \cb{X \in L^p \mid \varrho(X) \leq 0}
\end{equation}
is an acceptance set. Moreover, it holds $A = A_{\varrho_A}$ and $\varrho = \varrho_{A_\varrho}$.
\end{proposition}

The condition $A \cap \R\One \neq \emptyset$ means that there is an amount of cash which the financial agent accepts, and $(L^p\backslash A) - \R\One = L^p$ says that there is a limit to cash withdrawals starting from whatever position $X$. Both conditions together make sure that $\varrho_A$ never attains $-\infty$ as a value and that $\varrho_A(0) \in \R$. Everything else being straightforward, one can verify $A_{\varrho_A} \subseteq A$ as follows: If $X \in A_{\varrho_A}$, then $\varrho_A\of{X} = \inf\cb{s \in \R \mid X + s\One \in A} \leq 0$, and the very definition of the infimum implies the following: For each $\eps > 0$ there is $s_\eps \leq \eps$ such that $X + s_\eps \One \in A$. Using the fact that $A$ is monotone one obtains
\[
X + \eps \One = X + s_\eps \One + \of{\eps - s_\eps}\One \in A + \R_+\One \subseteq A.
\]
Since $A$ is directionally closed, $X \in A$, hence $A_{\varrho_A} \subseteq A$.

Directional closedness of $A$ implies the closedness of the set $\cb{s \in \R \mid X + s\One \in A}$ which means that the infimum in the definition of $\varrho_A(X)$ is attained if it is not $+\infty$: there is $s_0 \in \R$ such that $\varrho_A(X) = s_0$ and $X + s_0\One \in A$. This implies $X + \varrho_A(X)\One \in A$, i.e., $X$ can be made acceptable by depositing $\varrho_A(X)$ units of cash (or more).

In most cases, risk measures and acceptance sets have to satisfy further requirements. Again, property \eqref{EqCashAdd} provokes one-to-one correspondences between properties of risk measures and acceptance sets:

(a) A risk measure $\varrho$ is convex if, and only if, the ``induced" acceptance set $A_\varrho$ is convex. This implies that a risk measure is convex if, and only if, it is quasiconvex.

(b) $\varrho$ is positively homogeneous (sublinear) if, and only if, $A_\varrho$ is a cone (a convex cone).

(c) $\varrho$ has (only) real values if, and only if, $A_\varrho - \R\One = L^p$.

Corresponding statements are obtained for acceptance sets $A$ and the ``induced" risk measures $\varrho_A$.  

Following \cite{ArtznerEtAl99MF}, it became custom in the math finance community to call sublinear risk measures {\em coherent}. However, the authors of \cite{ArtznerEtAl99MF} probably intended to use the word ``coherent" in a more literal sense: for example in \cite{HeathKu04MF} convex, but not necessarily sublinear risk measures are also called (weakly) coherent. Moreover, ``coherent" is also used in the sense which is usually associated with ``arbitrage-free"--as in \cite{Walley91}. 

A few elementary examples for risk measures are the following. The function $X \mapsto E\sqb{-X}$ is a linear risk measure on $L^1$, the function $X \mapsto -\essinf X$ is a sublinear risk measure on $L^\infty$. A large class of risk measures is based on quantiles: For $\alpha \in (0,1]$, the number
\[
q^+_\alpha(X) = \inf\cb{t \in \R \mid P[X \leq t] > \alpha} = \sup\cb{t \in \R \mid P[X < t] \leq \alpha}
\]
is called the upper $\alpha$-quantile of $X$; the function
\[
X \mapsto V@R_\alpha(X) = -q^+_\alpha(X) = \inf\cb{t \in \R \mid P[X + t\One < 0] \leq \alpha}
\]
is called the Value-at-Risk of $X$ at level $\alpha$; the function
\[
X \mapsto AV@R_\alpha(X) = \frac{1}{\alpha}\int_0^\alpha V@R_\beta(X) d\beta
\] 
is called the Average-Value-at-Risk of  $X$ at level $\alpha$. Whereas $V@R$ is a positively homogeneous, but in general non-convex risk measure on $L^0$, the $AV@R$ is a sublinear risk measure on $L^1$, i.e., it is coherent.

The function $\tau \colon L^p \to \R\cup\cb{+\infty}$ defined by
\[
\tau(X) = 
	\left\{
	\begin{array}{ccc}
	-r & \colon & X = r\One, \; r \in \R \\
	+\infty & \colon & X \; \text{is non-constant}
	\end{array}
	\right.
\]
is sublinear and satisfies the requirements of Definition \ref{DefRiskMeasure} except for monotonicity. It could be seen as an extreme way to evaluate risk: non-constant payoffs are considered as intolerable risks, and acceptable are only the non-negative constant ones.
It turns out that every risk measure has a representation in terms of $\tau$. Indeed, defining the indicator function (in the sense of convex analysis) $I_A \colon L^p \to \R\cup\cb{+\infty}$ of an acceptance set $A \subseteq L^p$ by $I_A(X) = 0$ whenever $X \in A$ and $I_A(X) = +\infty$ otherwise, formula \eqref{EqRhoA} can be written as
\begin{align*}
\of{I_A \boxempty \tau}(X) & = \inf\cb{I_A(X_1) + \tau(X_2) \mid X_1 + X_2 = X} \\
 	& = \inf\cb{I_A(X_1) -r \mid X_1 + r\One = X, r \in \R}  \\
	& =  \inf\cb{-r \mid X - r\One \in A, \, r \in \R} = \varrho_A(X).
\end{align*}
Thus, the position $X$ is split into a constant and a remaining position $X_1$ which should be acceptable. If this is possible, one looks for the minimal risk of the constant evaluated by $\tau$. If such a split is not possible, $I_A(X_1) = +\infty$ always holds and $\of{I_A \boxempty \tau}(X) = +\infty$. Mathematically, $\varrho_A$ is the infimal convolution of $I_A$ and $\tau$. Every risk measure that satisfies the assumptions in Proposition \ref{PropPrimalRep} has such a representation: $\varrho(X) =  \of{I_{A_\varrho} \boxempty \tau}(X)$ for all $X \in L^p$. This is very convenient, in particular for duality purposes, since the two functions $I_A$ and $\tau$ are easy to handle. 

%%%New section
\section{Closedness and dual representation} 

The space $L^0$ is a complete metric, linear space for any L\'evy-metric. If $p \geq 1$, $L^p$ is a Banach space with the norm $\norm{X}_p = \of{\int_\Omega \abs{X}^pdP}^\frac{1}{p}$ for $1 \leq p < \infty$ and $\norm{X}_\infty = \esssup \abs{X}$ for $p = \infty$. In the following, $p \in \cb{0} \cup [1, \infty]$ is assumed.

\begin{proposition}
\label{PropClosedness} The following statements are equivalent for a risk measure $\varrho \colon L^p \to \R\cup\cb{+\infty}$:

(a)  At each $X \in L^p$, $\varrho$ is lower semicontinuous, i.e., $\varrho(X) \leq \liminf_{n \to \infty}  \varrho(X_n)$ whenever $\lim_{n \to \infty} X_n = X$ in $L^p$.

(b) $\cb{X \in L^p \mid \varrho(X) \leq r}$ is closed for each $r \in \R$;

(c) $A_\varrho = \cb{X \in L^p \mid \varrho(X) \leq 0} \subseteq L^p$ is closed.

A parallel statement holds for $\varrho_A$ replaced by a risk measure $\varrho$ and $A$ by $A_\varrho$.
\end{proposition}

The equivalence of (a) and (b) is standard in variational analysis, while (c) enters the picture because of the cash-additivity \eqref{EqCashAdd}. A risk measure that satisfies one (and hence) all of the conditions in Proposition \ref{PropClosedness} is called {\em closed}.
 
For $p \in [1, \infty)$, the topological dual of $L^p$ is the Banach space $L^q$ for $\frac{1}{p} + \frac{1}{q} = 1$ with $q = \infty$ whenever $p = 1$. If $p = \infty$ then $L^\infty$ is supplied with the weak topology generated by the dual pair $(L^1, L^\infty)$ (of locally convex spaces, see \cite[Section 5.14]{AliprantisBorder06}), and this ensures that $L^\infty$ and $L^1$ become topological duals of each other. Note that the topology on $L^\infty$ influences the closedness of $\varrho$: There are functions on $L^\infty$ which are closed with respect to the norm topology, but not closed with respect to the (weak) topology generated by $L^1$. A condition that ensures the latter turns out to be equivalent to the so-called Fatou property, see \cite[Section 4.3]{FoellmerSchied11}.

Let $\varrho \colon L^p \to \R\cup\cb{+\infty}$ be a closed, convex risk measure. According to Definition \ref{DefRiskMeasure}, it never attains the value $-\infty$, and it has at least one real value $\varrho(0)$. This means that $\varrho$ is proper in the sense of convex analysis (see \cite[p. 39]{Zalinescu02}), and it satisfies all the assumptions of the Fenchel-Moreau theorem \cite[Theorem 2.3.3]{Zalinescu02}: it coincides with its Legendre-Fenchel biconjugate $\varrho^{**}$ which is given by the two formulas
\[
\varrho^*\of{Y} := \sup_{X \in L^p}\cb{E\sqb{XY} - \varrho(X)} \quad \text{and} \quad
\varrho^{**}\of{X} := \sup_{Y \in L^q}\cb{E\sqb{XY} - \varrho^*(Y)}
\]
where the Legendre-Fenchel conjugate $\varrho^* \colon L^\infty \to \R\cup\cb{+\infty}$ of $\varrho$ is defined on the topological dual space $L^q$ of $L^p$. 

The representation $\varrho = \varrho^{**}$ is useful only if one can determine $\varrho^*$. It turns out that 
\begin{equation}
\label{EqConjugateRM}
\varrho^*(Y) = 
\left\{
\begin{array}{ccc}
\sup_{X \in A_\varrho} E\sqb{-XY} & : & E\sqb{Y} = 1, \; Y \in L^\infty_+ \\
+\infty   & : & \text{otherwise}
\end{array}
\right.
\end{equation}
This follows from the representation $\varrho = I_{A_\varrho} \boxempty \tau$ and the fact that the conjugate of an infimal convolution is the sum of the conjugates (\cite[Theorem 2.3.1 (ix)]{Zalinescu02}): one has to compute $(I_{A_\varrho})^*$ and $\tau^*$. While the former is known to be the support function of $A_\varrho$ (an easy consequence of the definition of the conjugate), the latter is $I_{\cb{Y \in L^q \mid E[Y] = -1}}$. Observing that the support function of $A_\varrho$ attains the value $+\infty$ whenever $Y \not\in -L^q_+$ (this follows from monotonicity of $A_\varrho$) and then replacing $Y$ by $-Y$ one obtains \eqref{EqConjugateRM}.

The two conditions for $Y$ in \eqref{EqConjugateRM} admit a striking interpretation of the dual representation formula $\varrho = \varrho^{**}$. To $Y \in L^q_+$ satisfying $E\sqb{Y} = 1$ one can assign a probability measure $Q$ by
\[
Q(A) = \int_A Y(\omega)dP \quad \text{for} \quad A \in \F
\]
which is absolutely continuous with respect to $P$, i.e., $\frac{dQ}{dP} = Y$. Moreover, the relationship between $Q$ and $Y$ is one-to-one. If one denotes the set of such probability measures by $\mathcal M_1(P)$, then the dual representation result for risk measures on $L^p$ reads as follows.

\begin{theorem}
\label{ThmDualRep} The function $\varrho \colon L^p \to \R\cup\cb{+\infty}$ is a closed, convex risk measure if, and only if, there exists a non-empty set $\mathcal Q_\varrho \subseteq \mathcal M_1(P)$ and a function $\gamma \colon \mathcal Q_\varrho \to \R$ such that
\[
\forall X \in L^p \colon \varrho(X) = \sup_{Q \in \mathcal Q_\varrho}\cb{E^Q\sqb{-X} - \gamma\of{Q}}.
\]
Moreover, $\gamma(Q) = \sup_{X \in A_\varrho} E\sqb{-XY} = \sup_{X \in A_\varrho} E^Q\sqb{-X}$ whenever $Q$ is the probability measure generated by $Y \in L^q_+$ which satisfies  $E\sqb{Y} = 1$ and $\varrho^*(Y) < +\infty$.

If $\varrho$ is additionally positive homogeneous (hence sublinear), then $\gamma(Q) = 0$ for $Q \in \mathcal Q_\varrho$.
\end{theorem} 

The worst case risk measure $\varrho_{max} \colon L^\infty \to \R$ defined by $\varrho_{max}(X) = - \essinf X = \inf\cb{t \in \R \mid X + t\One \geq 0}$ has the dual representation
\[
\varrho_{max}(X) = \sup_{Q \in \mathcal M_1(P)} E^Q[-X],
\]
whereas the Average-Value-at-Risk on $L^1$ can be represented as
\[
AV@R_\alpha(X) = \sup\cb{E^Q[-X] \mid Q \in \mathcal M_1(P), \;  \frac{dQ}{dP} \leq \frac{1}{\alpha}}.
\]
Both are sublinear (coherent) risk measures. Note that $\frac{dQ}{dP} \in L^\infty$ for $Q \in \mathcal Q_{AV@R}$. A verification of this formula can be given via the representation $AV@R_\alpha(X) = (\varphi \boxempty \tau)(X)$ with $\varphi(X) = \frac{1}{\alpha}\max\cb{-X, 0}$ which is due to \cite{RockafellarUryasev00JR}, \cite{RockafellarUryasev02JBF}.
 
By $\varrho_{ent}(X) = \frac{1}{\beta}\log E[exp(-\beta X)]$ for $\beta > 0$, a risk measure $\varrho_{ent} \colon L^\infty \to \R$ is defined; it is convex, but not positively homogeneous. Its dual representation is 
\[
\varrho_{ent}(X) = \sup_{Q \in \mathcal M_1(P)} \cb{E^Q[-X] - \frac{1}{\beta}H(Q\mid P)}
\] 
where $H(Q\mid P) = E^Q[\log \frac{dQ}{dP}]$ is the relative entropy of $Q$ with respect to $P$.

%%%New section
\section{Law invariance and Kusuoka representation} 

A risk measure $\varrho \colon L^p \to \R\cup\cb{+\infty}$ is called {\em law invariant} if $\varrho(X) = \varrho(Y)$ whenever $X$ and $Y$ have the same distribution under $P$. Standard examples of law invariant risk measures are the quantile based $V@R$ and $AV@R$. For risk measures on $L^\infty$, law invariance has strong implications. A typical result reads as follows.

\begin{theorem}
\label{ThmKusuoka}
Let $(\Omega, \F, P)$ be an atomless probability space such that $L^2$ is separable. Then, $\varrho \colon L^\infty \to \R$ is a law invariant convex risk measure if, and only if, there exists a convex function $\pi \colon \mathcal M_1((0,1]) \to [0, \infty]$ such that
\[
\forall X \in L^\infty \colon \varrho(X) = \sup_{m \in \mathcal M_1((0,1])}\cb{\int_0^1 AV@R_\alpha(X)dm(\alpha) - \pi(m)}.
\]
where $\mathcal M_1((0,1])$ is the set of (Borel) probability measures on $(0,1]$.
\end{theorem}

The characterization in Theorem \ref{ThmKusuoka} is due to Kusuoka \cite{Kusuoka01} for the sublinear case (in terms of integrated quantile functions) and due to Jouini, Schachermayer  and Touzi \cite{JouiniSchachermayerTouzi06} in the general convex case. It shows the importance of the Average-Value-at-Risk.

%%%New section
\section{Constructing risk measures}

{\bf Translative envelopes.} Let $\phi \colon L^1 \to \R\cup\cb{+\infty}$ be a monotone function. Then, the function $\varrho_\phi \colon L^1 \to \R\cup\cb{+\infty}$ defined by
\begin{equation}
\label{EqInfConvRM}
\varrho_\phi(X) = \inf\cb{\phi(X_1) +  \tau\of{X_2} \mid X_1 + X_2 = X}  
	 = \inf\cb{\phi\of{X - r\One} - r \mid r \in \R}
\end{equation}
is a risk measure whenever $\varrho_\phi(0) \in \R$. Note that $\varrho_\phi$ is nothing else than the infimal convolution of the two functions $\phi$ and $\tau$ (\cite[Theorem 2.1.3 (ix)]{Zalinescu02}). Moreover, it can be shown that  $\varrho_\phi$ is the pointwise greatest cash-additive function which is pointwise not greater than $\phi$, thus it may be called the (lower) cash-additive envelope of $\phi$. This construction has been introduced in \cite{DoleckiGreco95TMNA} in a different context, and for risk measures in \cite{FilipovicKupper07IME}. Moreover, the so-called ``optimized certainty equivalent" introduced in \cite{BenTalTeboulle86MS}, \cite{BenTalTeboulle07MF} has the same form in which $\phi(X) = E[\ell(X)]$ for a monotone (non-increasing) function $\ell \colon \R \to \R\cup\cb{+\infty}$. As shown above, every risk measure is the cash-additive envelope of the indicator function of its ``induced" acceptance set: $I_{A_\varrho}$ is monotone since $A_\varrho$ is.

\medskip {\bf Risk measures associated with loss/utility functions.} Let $\ell \colon \R \to \R\cup\cb{+\infty}$ be a proper, increasing and not identically constant function and $r_0 \in \Int \ell(\R)$. Define the set $A_\ell = \cb{X \in L^1 \mid E\sqb{\ell(-X)} \leq r_0}$. The risk measure $\varrho_\ell$ defined by
\[
\varrho_\ell(X) = \varrho_{A_\ell}(X) = \inf\cb{s \in \R \mid E\sqb{\ell(-X - s\One)} \leq r_0}
\]
is called {\em loss-based shortfall risk measure.} It is convex if $\ell$ is convex. If $\ell$ is real-valued and $\varrho_\ell$ is considered as a function on $L^\infty$, then it is weakly closed with dual representation
\[
\forall X \in L^\infty \colon \varrho_\ell(X) = \max_{Q \in \mathcal M_1(P)}\sqb{E^Q[-X] - \inf_{\lambda>0}\frac{1}{\lambda}\of{r_0 + E\sqb{\ell^*\of{\lambda\frac{dQ}{dP}}}}}
\]
where $\ell^*$ is the Fenchel conjugate of $\ell \colon \R \to \R$. Shortfall risk measures are law invariant and in some sense dual to {\em divergence risk measures} (discussed in \cite[Section 4.9]{FoellmerSchied11}, the latter have a primal representation depending on $\ell^*$) which in turn also coincide with the ``optimized certainty equivalent" introduced by Ben-Tal and Teboulle \cite{BenTalTeboulle86MS}, \cite{BenTalTeboulle07MF}.

\medskip {\bf Spectral risk measures.} The crucial observation is that a convex combination of two risk measures again is a risk measure, and this can even be generalized to mixtures via probability measures on $[0,1]$, see \cite[Proposition 2.2]{Acerbi02JBF}. Acerbi \cite{Acerbi02JBF} introduced the following concept. Let $\phi \colon [0,1] \to \R$ be a function satisfying (a) $\phi(\alpha) \geq 0$ for all $\alpha \in [0,1]$, (b) $\int_0^1 \phi(\alpha)d\alpha = 1$, (c) $0 \leq \alpha_1 \leq \alpha_2 \leq 1$ implies $\phi(\alpha_1) \geq \phi(\alpha_2)$. Then, the function $\varrho_\phi \colon L^\infty \to \R\cup\cb{+\infty}$ defined by
\[
\varrho_\phi(X) = -\int_0^1 \phi(s) q^-_X(s)ds
\]
is a coherent, law invariant risk measure, and the function $\phi$ is called a {\em risk spectrum} which can chosen by the decision maker. Here, $q^-_X(\alpha) = \inf\cb{t \in \R \mid F_X(t) \geq \alpha}$ is the {\em lower $\alpha$-quantile} of $X$. $V@R$ and $AV@R$ turn out to be special {\em spectral risk measures}. Compare \cite{CheriditoLi08MFE} for further properties, dual representation results and relationships to stochastic dominance orders. Note that already the results of Kusuoka \cite[Theorem 7]{Kusuoka01} imply that the class of spectral risk measures on $L^\infty$ over an atomless probability space coincides with the class of all weakly closed, coherent, law invariant and comonotonic risk measures (compare Remark 4.4 in \cite{Acerbi02JBF}).

%%%New section
\section{Relationships to other concepts in risk evaluation} 

{\bf Stochastic dominance orders.} Stochastic dominance orders for probability distributions are important tools for risk evaluation. Therefore, a crucial property of a risk measure is monotonicity with respect to these orders. The Average Value at Risk does even characterize the second order stochastic dominance $\preceq_{SSD}$: If $X, Y \in L^1$, then
\[
X \preceq_{SSD} Y \quad \Leftrightarrow \quad \forall \alpha \in (0,1] \colon AV@R_\alpha(X) \geq AV@R_\alpha(Y).
\] 
This observation goes back to \cite{OgryczakRuszczinski02SITOR}, see also \cite[Remark 4.49]{FoellmerSchied11}. In a similar way, the Value-at-Risk characterize first order stochastic dominance.

\medskip {\bf Other translative functions.} Remarkably, many other functions share property \eqref{EqCashAdd}. In particular, the sub- and superhedging price of a financial position in an incomplete market are versions of a cash-additive function \cite[Section 1.3]{FoellmerSchied11} and also the so-called {\em good deal bounds} \cite{JaschkeKuechler01FS}. Outside finance, Dempster's belief functions \cite[formula (3.9), p. 363]{Dempster66AMS}, Choquet integrals \cite{Denneberg94}, imprecise lower/upper expectations \cite{Walley91}, insurance premiums as discussed, e.g., in \cite{WangYoungPanjer97IME}, exact functionals and games \cite{Maass01}, \cite{Maass02} as well as maxmin expected utility functions \cite{GilboaSchmeidler89}, among many others, share property \eqref{EqCashAdd}.  

\medskip {\bf Extensions.} (a) The famous Markowitz model for portfolio selection \cite{Markowitz52JF} involves the variance as a risk evaluating tool - which is neither monotone, nor cash-additive. On the contrary, it is constant on the linear subspace of $L^2$ formed by the constant functions. This property is shared by {\em deviation measures} introduced by Rockafellar, Uryasev and Zabarankin \cite{RockafellarEtAl06FS}, \cite{RockafellarEtAl06MP} which are basically the difference of a risk measure and the expected value. They may replace the variance in procedures like regression analysis \cite{RockafellarEtAl08MOR} or portfolio selection \cite{RockafellarEtAl06MP}. See \cite{RockafellarUryasev13SOR} for an overview. (b) Since a cash-additive risk measure is quasiconvex if, and only if, it is convex, weaker versions of \eqref{EqCashAdd} were introduced, see \cite{ElKarouiRavanelli09MF} and \cite{CerreiaVoglioEtAl11MF}. In \cite{CernyMadan09RFS}, \cite{DrapeauKupper13MOR}, a concise motivation, further results on quasiconvex risk measures (called performance or assessment indices) and many examples can be found. (c) Under market conditions, one may want to make available a dynamic risk assessment procedure. The main issue is time-consistency, i.e., a position which is acceptable at some point in time should already be acceptable at earlier times. Extensions of the above concepts to the dynamic case were initiated in \cite{DetlefsenScandolo05FS}, \cite{CheriditoDelbaenKupper04SPA}, \cite{CheriditoDelbaenKupper05FS}, \cite{Riedel04SPA}. More recently, the $L^0$-module framework was developed mainly motivated by time-dependent, conditional risk measures, see  \cite{FilipovicKupperVogelpoth12SIAMF} for an overview and references. (d) In markets with transaction costs and illiquidity, the risk of multi-variate positions needs to be evaluated (see also {\f 3.1.5.7}). Several approaches have been pursued: scalar risk measures for multivariate payoffs \cite{BurgertRueschendorf06IME}, \cite{EkelandSchachermayer11SRM}, for example, and vector- and set-valued risk measures \cite{JouiniEtAl04FS}, \cite{HamelHeyde10SIAMFM}, \cite{HamelHeydeRudloff11MFE}, \cite{HamelKostner18JMVA}. (e) Condition \eqref{EqCashAdd} requires the existence of a ``non-defaultable" (discountable) num\'eraire which serves as reference instrument. In the light of recent financial and economic crises, this assumption is questionable. Even more reasons for leaving the framework of ``constant num\'eraires" and alternatives can be found in \cite{FarkasKochMedinaMunari14IME}, \cite{FarkasKochMedinaMunari14FS}.


\begin{thebibliography}{9999}

\bibitem{Acerbi02JBF} Acerbi, C. (2002). Spectral measures of risk: a coherent representation of subjective risk aversion. Journal of Banking \& Finance 26(7): 1505-1518.

\bibitem{AcerbiScandolo08QF} Acerbi, C., \& Scandolo ¤, G. (2008). Liquidity risk theory and coherent measures of risk. Quantitative Finance, 8(7): 681-692.

\bibitem{AliprantisBorder06} Aliprantis, C., \& Border, K. (2006). Infinite Dimensional Analysis. Springer Publishers, 3rd edition.

\bibitem{ArtznerEtAl99MF} Artzner, P., Delbaen, F., Eber, J. M., \& Heath, D. (1999). Coherent measures of risk. Mathematical Finance, 9(3), 203-228.

\bibitem{BenTalTeboulle86MS} Ben-Tal, A., \& Teboulle, M. (1986). Expected utility, penalty functions, and duality in stochastic nonlinear programming. Management Science, 32(11), 1445-1466.

\bibitem{BenTalTeboulle07MF} Ben-Tal, A., \& Teboulle, M. (2007). An old-new concept of convex risk measures: the optimized certainty equivalent. Mathematical Finance, 17(3), 449-476.

\bibitem{BurgertRueschendorf06IME} Burgert, C., \& R\"uschendorf, L. (2006). Consistent risk measures for portfolio vectors. Insurance: Mathematics and Economics, 38(2), 289-297.

\bibitem{CernyMadan09RFS} Cherny, A., \& Madan, D. (2009). New measures for performance evaluation. Review of Financial Studies, 22(7), 2571-2606.

\bibitem{CerreiaVoglioEtAl11MF} Cerreia--Vioglio, S., Maccheroni, F., Marinacci, M., \& Montrucchio, L. (2011). Risk measures: rationality and diversification. Mathematical Finance, 21(4), 743-774.

\bibitem{CheriditoDelbaenKupper04SPA} Cheridito, P., Delbaen, F., \& Kupper, M. (2004). Coherent and convex monetary risk measures for bounded cadlag processes. Stochastic Processes and their Applications, 112(1), 1-22.

\bibitem{CheriditoDelbaenKupper05FS} Cheridito, P., Delbaen, F., \& Kupper, M. (2005). Coherent and convex monetary risk measures for unbounded cadlag processes. Finance and Stochastics, 9(3), 369-387.

\bibitem{CheriditoLi08MFE} Cheridito, P., \& Li, T. (2008). Dual characterization of properties of risk measures on Orlicz hearts. Mathematics and Financial Economics, 2(1), 29-55.

\bibitem{Delbaen02} Delbaen, F. (2002). Coherent risk measures on general probability spaces. In: Advances in Finance and Stochastics (pp. 1-37). Springer Publishers.

\bibitem{Dempster66AMS} Dempster, A. P. (1966). New methods for reasoning towards posterior distributions based on sample data. The Annals of Mathematical Statistics, 355-374.

\bibitem{Denneberg94} Denneberg, D. (1994). Non-additive measure and integral. Kluwer Academic Publishers Dordrecht.

\bibitem{DetlefsenScandolo05FS} Detlefsen, K., \& Scandolo, G. (2005). Conditional and dynamic convex risk measures. Finance and Stochastics, 9(4), 539-561.

\bibitem{DoleckiGreco95TMNA} Dolecki, S., \& Greco, G. H. (1995). Niveloids. Topological Methods in Nonlinear Analysis, 5(1), 1-22.

\bibitem{DrapeauKupper13MOR} Drapeau, S., \& Kupper, M. (2013). Risk preferences and their robust representation. Mathematics of Operations Research, 38(1), 28-62.

\bibitem{ElKarouiRavanelli09MF} El Karoui, N., \& Ravanelli, C. (2009). Cash subadditive risk measures and interest rate ambiguity. Mathematical Finance, 19(4), 561-590.

\bibitem{EkelandSchachermayer11SRM} Ekeland, I., \& Schachermayer, W. (2011). Law invariant risk measures on $L^\infty(\R^d)$. Statistics \& Risk Modeling with Applications in Finance and Insurance, 28(3), 195-225.

\bibitem{FarkasKochMedinaMunari14IME} Farkas, W., Koch-Medina, P. \& Munari, C. (2014). Capital requirements with defaultable securities. Insurance: Mathematics and Economics 55: 58-67.

\bibitem{FarkasKochMedinaMunari14FS} Farkas, W., Koch-Medina, P. \& Munari, C. (2014). Beyond cash-additive risk measures: when changing the numŽraire fails. Finance \& Stochastics 18(1): 145-173.

\bibitem{FilipovicKupper07IME} Filipovi\'{c}, D., \& Kupper, M. (2007). Monotone and cash-invariant convex functions and hulls. Insurance: Mathematics and Economics, 41(1), 1-16.

\bibitem{FilipovicKupperVogelpoth12SIAMF} Filipovi\'{c}, D., Kupper, M., \& Vogelpoth, N. (2012). Approaches to conditional risk. SIAM Journal on Financial Mathematics, 3(1), 402-432.

\bibitem{FoellmerSchied02FS} F\"ollmer, H., \& Schied, A. (2002). Convex measures of risk and trading constraints. Finance and Stochastics, 6(4), 429-447.

\bibitem{FoellmerSchied02} F\"ollmer, H., \& Schied, A. (2002). Robust preferences and convex measures of risk. In Advances in Finance and Stochastics (pp. 39-56). Springer Publishers.

\bibitem{FoellmerSchied11} F\"ollmer, H., \& Schied, A. (2011). Stochastic Finance: An Introduction in Discrete Time. Walter de Gruyter, 3rd edition.

\bibitem{FritelliRosazza02JBF}  Frittelli, M., \& Rosazza Gianin, E. (2002). Putting order in risk measures. Journal of Banking \& Finance, 26(7), 1473-1486.

\bibitem{GilboaSchmeidler89} Gilboa, I., \& Schmeidler, D. (1989). Maxmin expected utility with non-unique prior. J. Mathematical Economics, 18(2), 141-153.

\bibitem{HamelHeyde10SIAMFM} Hamel, A. H., \& Heyde, F. (2010). Duality for set-valued measures of risk. SIAM J. Financial Mathematics, 1(1):66-95.

\bibitem{HamelHeydeRudloff11MFE} Hamel, A. H., Heyde, F., \& Rudloff, B. (2011). Set-valued risk measures for conical market models. Mathematics and Financial Economics, 5(1):1-28.

\bibitem{HamelKostner18JMVA} Hamel, A. H., \& Kostner, D. (2018). Cone distribution functions and quantiles for multivariate random variables, J. Multivariate Analysis 167, 97-113.

\bibitem{HeathKu04MF} Heath, D. \& Ku, H. (2004). Pareto equilibria with coherent measures of risk. Math. Finance, 14(2):163Ð172.

\bibitem{JaschkeKuechler01FS} Jaschke, S., \& K\"uchler, U. (2001). Coherent risk measures and good-deal bounds. Finance and Stochastics, 5(2), 181-200.

\bibitem{JouiniEtAl04FS} Jouini, E., Meddeb, M., \& Touzi, N. (2004). Vector-valued coherent risk measures. Finance and Stochastics, 8(4), 531-552.

\bibitem{JouiniSchachermayerTouzi06} Jouini, E., Schachermayer, W., \& Touzi, N. (2006). Law invariant risk measures have the Fatou property. In Advances in Mathematical Economics (pp. 49-71). Springer Publishers.

\bibitem{Kusuoka01} Kusuoka, S. (2001). On law invariant coherent risk measures. In: Advances in Mathematical Economics (pp. 83-95). Springer Publishers.

\bibitem{Markowitz52JF} Markowitz, H. (1952). Portfolio selection. The Journal of Finance, 7(1), 77-91.

\bibitem{Maass01} Maa{\ss}, S. (2001). Coherent Lower Previsions as Exact Functionals and their (Sigma-)Core. In ISIPTA, Vol. 1, pp. 230-236.

\bibitem{Maass02} Maa{\ss}, S. (2002). Exact functionals and their core. Statistical Papers, 43(1), 75-93.


\bibitem{OgryczakRuszczinski99EJOR} Ogryczak, W., \& Ruszczynski, A. (1999). From stochastic dominance to mean-risk models: Semideviations as risk measures. European J. Operational Research, 116(1), 33-50.

\bibitem{OgryczakRuszczinski02SITOR} Ogryczak, W., \& Ruszczynski, A. (2002). Dual stochastic dominance and quantile risk measures. International Transactions in Operational Research, 9(5), 661-680.

%\bibitem{OgryczakRuszczinski02SIAMJO} Ogryczak, W., \& Ruszczynski, A. (2002). Dual stochastic dominance and related mean-risk models. %SIAM Journal on Optimization, 13(1), 60-78.

\bibitem{Riedel04SPA} Riedel, F. (2004). Dynamic coherent risk measures. Stochastic processes and their applications, 112(2), 185-200.

\bibitem{RockafellarUryasev00JR} Rockafellar, R. T., \& Uryasev, S. (2000). Optimization of conditional value-at-risk. Journal of Risk, 2, 21-42.

\bibitem{RockafellarUryasev02JBF} Rockafellar, R. T., \& Uryasev, S. (2002). Conditional value-at-risk for general loss distributions. Journal of Banking \& Finance, 26(7), 1443-1471.

\bibitem{RockafellarUryasev13SOR} Rockafellar, R. T., \& Uryasev, S. (2013). The fundamental risk quadrangle in risk management, optimization and statistical estimation. Surveys in Operations Research and Management Science, 18(1), 33-53.

\bibitem{RockafellarEtAl06FS} Rockafellar, R. T., Uryasev, S., \& Zabarankin, M. (2006). Generalized deviations in risk analysis. Finance and Stochastics, 10(1), 51-74.

\bibitem{RockafellarEtAl06MP} Rockafellar, R. T., Uryasev, S., \& Zabarankin, M. (2006). Optimality conditions in portfolio analysis with general deviation measures. Mathematical Programming, 108(2-3), 515-540.

\bibitem{RockafellarEtAl08MOR} Rockafellar, R. T., Uryasev, S., \& Zabarankin, M. (2008). Risk tuning with generalized linear regression. Mathematics of Operations Research, 33(3), 712-729.

\bibitem{RoordaSchumacherEngwerda05MF} Roorda, B., Schumacher, J. M., \& Engwerda, J. (2005). Coherent acceptability measures in multiperiod models. Mathematical Finance, 15(4), 589-612.

\bibitem{RuszczynskiShapiro06MOR} Ruszczynski, A., \& Shapiro, A. (2006). Optimization of convex risk functions. Mathematics of Operations Research, 31(3), 433-452.

\bibitem{Walley91} Walley, P. (1991). Statistical Reasoning with Imprecise Probabilities. London: Chapman and Hall.

\bibitem{WangYoungPanjer97IME} Wang, S. S., Young, V. R., \& Panjer, H. H. (1997). Axiomatic characterization of insurance prices. Insurance: Mathematics and Economics, 21(2), 173-183.

\bibitem{Yaari87E} Yaari, M. E. (1987). The dual theory of choice under risk. Econometrica, 95-115.

\bibitem{Zalinescu02} Z\u{a}linescu, C. (2002). Convex Analysis in General Vector Spaces.
World Scientific Singapore.

\end{thebibliography}
\end{document}